\documentclass{article}
\usepackage[final]{neurips}

\usepackage{graphicx}% Include figure files
\usepackage{dcolumn}% Align table columns on decimal point
\usepackage{bm}% bold math
\usepackage{hyperref}% add hypertext capabilities
\usepackage{bbm}
\usepackage{physics}
\usepackage{float}
\usepackage{booktabs}

\title{An Empirical Review of Optimization Techniques for Quantum Variational Circuits}% Force line breaks with \\
\author{Owen Lockwood \\
Department of Computer Science \\
 Rensselaer Polytechnic Institute, Troy NY, USA \\
  \texttt{lockwo@rpi.edu}
}%

\begin{document}

\maketitle

\begin{abstract}
Quantum Variational Circuits (QVCs) are often claimed as one of the most potent uses of both near term and long term quantum hardware. The standard approaches to optimizing these circuits rely on a classical system to compute the new parameters at every optimization step. However, this process can be extremely challenging, due to the nature of navigating the exponentially scaling complex Hilbert space, barren plateaus, and the noise present in all foreseeable quantum hardware. Although a variety of optimization algorithms are employed in practice, there is often a lack of theoretical or empirical motivations for this choice. To this end we empirically evaluate the potential of many common gradient and gradient free optimizers on a variety of optimization tasks. These tasks include both classical and quantum data based optimization routines. Our evaluations were conducted in both noise free and noisy simulations. The large number of problems and optimizers evaluated yields strong empirical guidance for choosing optimizers for QVCs that is currently lacking.

%\begin{description}
%\end{description}
\end{abstract}


%\tableofcontents

\section{\label{sec:intro}Introduction} 
%}First-level heading:\protect\\ The line break was forced \lowercase{via} \textbackslash\textbackslash}

Although theoretically proposed many decades ago \citep{feynman2018simulating}, only in recent years has the spectre of quantum computers become something of a reality \citep{arute2019quantum, zhong2020quantum, wu2021strong, arrazola2021quantum}. In this era of Noisy Intermediate Scale Quantum (NISQ) \citep{preskill2018quantum} devices, the number and quality of qubits are lacking to employ the foundational quantum algorithms (e.g. Shor's algorithm \citep{shor1994algorithms, gidney2021factor}). To this end, a number of Quantum Machine Learning (QML) routines have been proposed \citep{cerezo2021variational, bharti2021noisy} for supervised \citep{rebentrost2014quantum, cong2019quantum, blank2020quantum}, unsupervised \citep{aimeur2007quantum, lloyd2014quantum, dallaire2018quantum}, and reinforcement learning \citep{chen2020variational, lockwood2020reinforcement, lockwood2021playing, skolik2021quantum, jerbi2021parametrized}. Although many of these algorithms have claims of exponential scaling properties that enable them to exploit smaller hardware systems in theory, there are still a number of challenges in practice. Optimization of these algorithms suffers from the barren plateaus problem that negatively impacts both gradient and gradient free optimizers \citep{mcclean2018barren, arrasmith2021effect}. Additionally, the optimization of these models are NP-Hard \citep{bittel2021training}. Finally, converting theoretical circuit structures to real hardware is extremely non-trivial \citep{nagarajan2021quantumcircuitopt}. However, even with these striking difficulties, progress has been made with claims of experimental demonstrations of advantages via quantum machine learning \citep{riste2017demonstration, liu2021rigorous, huang2021quantum}.

In the training routines of QVCs there are a number of methods employed to classically calculate the parameter updates, including traditional black block optimizers, parameter-shift based gradient optimizers \citep{wierichs2021general}, and machine learning based optimizers \citep{sordal2019deep, yao2020policy, lockwood2021optimizing}. There is often little theoretical motivation for any given optimizer, although recent work has been done to analyze the training dynamics of QVCs \citep{liu2021representation}. In absence of compelling theoretical results, we evaluate optimizers empirically to determine how to best optimize a QVC for a given problem. To this end, we evaluate different QML problems of a variety of sizes ranging from 2 qubit systems with $< 10$ parameters to 25 qubit systems with 200 parameters. These include systems with no noise, only shot noise, and shot and depolarizing noise. These empirical evaluations were conducted using TensorFlow-Quantum \citep{broughton2021tensorflow}. Our paper expands upon existing works \citep{sung2020using, lavrijsen2020classical, bonet2021performance} by increasing the scope of the problem settings (focusing on a variety of QML problems, not just QAOA or VQE), the size of the problems, and the number of optimizers (experimenting a total of 46 different optimizer setups). Our analysis focuses only on the final performance of the optimizer and not the scaling or shots required. As a result of these experiments we are able to offer the following empirically justified recommendations for optimizing quantum circuits: 

\begin{itemize}

    \item In general, parameter shift gradient optimizers outperform non parameter shift gradient based optimizers

    \item The best performing non-parameter shift gradient optimizers are Powell's \citep{powell1964efficient} and Simultaneous Perturbation Stochastic Approximation (SPSA) \citep{spall1998overview} 
    
    \item For parameter shift gradient optimizers, in the absence of the ability to perform hyperparameter sweeps, the best performing is Adam \citep{kingma2017adam} with learning rate 0.1
    
    \item For parameter shift gradient optimizers, with the ability to perform hyperparameter sweeps, evaluate Adam, AMSGrad \citep{reddi2019convergence}, and NAdam \citep{dozat} with learning rate $\in \{0.01, 0.1\}$
    
    \item For all cases, also run an evaluation Nelder-Mead \citep{nelder1965simplex}
    
\end{itemize}

These recommendations differ from common intuition for both classical neural networks (with the learning rates being far higher for the gradient based optimizers) and noisy optimization (with the emphasis on Nelder-Mead), but remain strongly grounded in empirical results. All code and results are available at: \href{https://github.com/lockwo/qvc_opt_review}{github.com/lockwo/qvc\_opt\_review}.

\section{\label{back}Background}

Quantum Machine Learning is a field lying at the intersection of quantum computing and machine learning. It seeks to use the advancements of quantum computing to accelerate and improve machine learning algorithms. Although we consider a variety of QML problems, they all share the same general formalization. In this QML problem setting the goal is to minimize the function $f(\theta) = C \left ( \langle U^\dagger(\theta) | \hat{M} | U(\theta) \rangle \right )$, where $\theta$ are the tunable parameters, $\hat{M}$ is the measurement operator, and $C$ is the cost function \citep{mari2021estimating}. $U(\theta)$ can be any unitary matrix parameterized by $\theta$ and is sufficiently general to capture any QML architecture, which necessitate unitary operations. It can also be comprised of any number of non-parameterized gates in addition to the parameterized gates. In order to differentiate this function, we use the parameter shift rule: $\nabla f(\theta) = \frac{f(\theta + s) - f(\theta - s)}{2 sin(s)}, s \in \mathbbm R, s \neq k \pi, k \in \mathbbm Z$ \citep{schuld2019evaluating}. We can then plug this analytic gradient into traditional gradient based optimizers (like those used in classical machine learning). We will now briefly review each of the optimizers used in this work. We select the following parameter shift gradient based optimizers: 

\begin{itemize}
    
    \item Stochastic Gradient Descent (SGD): at every iteration a step is taken in the direction of the negative gradient, the size of which is dictated by the learning rate
    
    \item Adaptive Gradient Algorithm (AdaGrad) \citep{duchi2011adaptive}: adaptively adjusts the learning rate according to $1/\sqrt{\epsilon I + diag \left (\sum_t g_t g_t^T \right ) }$, with the summation being over the previous gradients
    
    \item AdaDelta \citep{zeiler2012adadelta}: building upon AdaGrad, it accumulates the past gradients only within some window keeping tract of the decaying average of past gradients squared
    
    \item Root Mean Square Propagation (RMSProp) \citep{tieleman2012lecture}: attempts to solve AdaGrad's problem of learning rate diminishing by keeping a running average of the gradients and updating the parameters at a rate according to the root mean squared of this average
    
    \item Follow The Regularized Leader (FTRL) \citep{mcmahan2013ad}: updates parameters individually via the following (in the unregulated form): \begin{gather*}
    n_{i} = n_{i-1} + g_t^2 \\ \sigma = \left ( \sqrt{n_i} - \sqrt{n_{i-1}} \right )/\eta \\ z_{i} = z_{i - 1} + g_t - \sigma \theta \\ \theta = sign(z) * \eta / \sqrt{n_{i}} \end{gather*}
    
    \item Adam \citep{kingma2017adam}: the most popular choice of optimizer for classical neural networks, it adjusts the learning rate according to adaptive estimations of the first and second moments
    
    \item Adamax \citep{kingma2017adam}: a variant of Adam that uses the $L^\infty$ norm instead of the $L^2$ norm to scale the gradients
    
    \item NAdam \citep{dozat}: a variant of Adam that adjusts its usage of momentum to be Nesterov momentum
    
    \item AMSGrad \citep{reddi2019convergence}: a variant of Adam that (incorrectly \citep{tran2019convergence}) fixed the proofs of convergence and adjusts the algorithm by keeping an updated maximum of gradients squared rather than the exponential averaging
    
\end{itemize}

We use the following non-parameter shift gradient based optimizers (some are gradient free and some simply use their own approximation of the gradients):

\begin{itemize}

\item Nelder-Mead \citep{nelder1965simplex}: simplex based method for direct minimization that required $O(1)$ forward passes for every update iteration. Along with SPSA, it is the only constant scaling optimizer 

\item Simultaneous Perturbation Stochastic Approximation (SPSA) \citep{spall1998overview}: performs gradient descent using a noisy estimate of the gradient defined by $\nabla f(\theta) = \frac{f(\theta + c  \Delta) - f(\theta - c \Delta)}{2c\Delta}$ where $\Delta$ is the perturbation vector (a collection of randomly chosen positive and negatives ones) 

\item Powell \citep{powell1964efficient}: takes $N$ search vectors and conducts bi-directional line searches on each vector, finds the minimum for each, updates these search vectors, and repeats this process

\item Nonlinear Conjugate Gradient (CG) \citep{fletcher1964function}

\item Broyden–Fletcher–Goldfarb–Shanno (BFGS) \citep{nocedal2006numerical}

\item Limited Memory Broyden–Fletcher–Goldfarb–Shanno (L-BFGS-B) \citep{byrd1995limited}

\item Newton Conjugate Gradient (TNC) \citep{nash1984newton}

\item Constrained Optimization By Linear Approximation (COBYLA) \citep{powell1994direct}

\item Sequential Least Squares Programming (SLSQP) \citep{kraft}

\item Byrd-Omojokun Trust-Region Sequential Quadratic Programming (trust-constr) \citep{lalee1998implementation}

\end{itemize}

All of these are implemented and provided by \cite{2020SciPy-NMeth} except SPSA which is done via noisyopt. We expand only a subset of these algorithms as they are the best performing and many of these algorithms have been around longer and are more established that the parameter shift gradient methods. These algorithms represent substantial coverage of standard black-box optimizers and gradients optimizers common in classical machine learning (especially neural networks).

\section{\label{methods}Methods}

To evaluate the optimizers we use a variety of problems and routines. Although there isn't a standard set of benchmarks for QML like there is in other fields (e.g. the Atari benchmark for RL), we use established quantum routines and previously used classical data problems. The problems/routines used as a benchmark suite for the optimizers in this work are:

\begin{itemize}

    \item Variational Quantum Eigensolver (VQE) \citep{peruzzo2014variational}: an optimization routine used to find the minimal eigenvalue of a given hamiltonian, formalized as $\min_\theta \langle \Psi (\theta) | \hat{H} | \Psi(\theta) \rangle$ where $\Psi (\theta)$ is the parameterized quantum circuit and $\hat{H}$ is the hamiltonian
    
    \item Quantum Approximate Optimization Algorithm (QAOA) \citep{farhi2014quantum}: an optimization routine for combinatorial optimization problems which utilizes alternating layers of cost and mixer hamiltonians, formalized as $\min_{\beta, \gamma} \langle \Psi (\beta, \gamma) | \hat{C} | \Psi (\beta, \gamma) \rangle$, where $\hat{C}$ is the cost function and $\Psi(\beta, \gamma) = e^{-i \beta_p H_m}e^{-i \gamma_p H_c} \hdots e^{-i \beta_0 H_m} e^{-i \gamma_0 H_c}|\Psi_0 \rangle$
    
    \item Excited state classification: a dataset of cluster states to classify as either containing a X gate or not, uses the mean squared error between the expected value of the Z operator on the last qubit and the target label as the loss function
    
    \item Moon binary classification: a generated dataset of two classes of two dimensional datapoints (moon, circle and blob visualizations can be seen in \citep{lockwood2021optimizing}), uses the cross entropy between the target label and the the expected values of the Z operator on both qubits as the loss function
    
    \item Circle binary classification: a generated dataset of two circles (one enclosed by the other) with two dimensional inputs, uses the cross entropy between the target label and the the expected value of the Z operator on first qubit as the loss function
    
    \item Blobs multiclass classification: a generated dataset of a collection of blobs with two dimensional inputs, uses the categorical cross entropy between the Z expectation values on all the qubits and the labels as the loss function
    
    \item Regression on the Boston Housing dataset: attempts to predict the cost of houses based on a 13 dimensional input, uses the mean squared error between the labels and the Z expectation value of the first qubit as the loss function
    
\end{itemize}

For each of these problems, we experiment with a set of variations (e.g. depth, depolarizing vs just shot noise, number of qubits). All variations and the numbers of parameters can be found in Table \ref{tab:exp1}. For the noisy simulations, the noise is modelled by a depolarizing channel after every gate, defined as $\rho \rightarrow \left ( 1 - p \right ) \rho + \frac{p}{4^n - 1} \sum_i P_i \rho P_i $, with $p = 0.01$. The ansatz of each circuit is focused on hardware efficiency with most being similar combinations of single qubit rotations and entangling gates. The exception is the excited state classifier, which utilizes a QCNN \citep{cong2019quantum} based architecture. All simulations had the following shared hyperparamters: maximum of 1000 optimizer iterations, convergence tolerance of $1 * 10^{-4}$, 3 repetitions of each problem with different random initializations, for each parameter shift gradient optimizer learning rate $\in \{0.0001, 0.001, 0.01, 0.1\}$, and 1000 shots to estimate the expectation value. The QAOA combinatorial problem is MAX-CUT for random 4-regular graphs. The hamiltonians for VQE are randomly generated and are composed of 10 terms (for all qubit sizes). For the large VQE simulations, all of the above parameters were the same with the exception of the number of shots. Due to the computational expense of large simulations, they were conducted in absence of shot noise, which allowed for efficient calculation of gradients using adjoint differentiation. Hence, only the gradient based optimizers where evaluated on these large VQE simulations. All other simulations (except the 20 and 25 qubit VQE) contained shot noise. 

\section{\label{res}Results}

There are two main metrics that we use to compare the optimizers: rank and normalized loss difference. The rank is simple, for each task we rank the optimizers (from best performing to least) and then average the rank over all tasks to get the final results. The loss difference calculates the following quantity for each task that is then averaged for the final results: $\frac{|loss_{opt} - \min loss|}{|\min loss|}$. This is a linearly decreasing function with the domain defined as $dom(f) = [\min loss, \infty]$.  The results shown in this section take the best performing learning rate of the parameter shift gradient based optimizers for each task. The full results (with each learning rate separated out) can be found in Appendix \ref{add}. In Table \ref{t:tank} we can see the (surprising) result that on average, Nelder-Mead ranks the highest of all optimizers. This is followed by Powell's method and several Adam variants. However, when we look at the average loss difference, shown in Table \ref{t:loss}, we can see results more in line with common understandings. Adam performs the best, with Adam variants close behind. This is indicative of the fact that Nelder-Mead is inconsistent in it's performance across tasks, but can perform well.

%\begin{figure}[h]
%    \centering
%    \hspace*{-2cm}
%    \includegraphics[scale=0.4]{rank.png}
%    \caption{Average Rank of Optimizers}
%    \label{fig:rank}
%\end{figure}
\begin{table}[h]
\centering
\begin{tabular}{|l|c|}
\hline
Optimizers & Average Rank  \\
\hline
Nelder-Mead & 4.4483 \\
\hline
Nadam & 4.8276 \\
\hline
Powell & 4.931 \\
\hline
AMSGrad & 5.7241 \\
\hline
Adam & 5.8621 \\
\hline
COBYLA & 6.1724 \\
\hline
RMSProp & 6.8966 \\
\hline
Adamax & 7.5862 \\
\hline
Ftrl & 7.5862 \\
\hline
Adagrad & 7.7586 \\
\hline
SPSA & 9.4483 \\
\hline
SGD & 10.5862 \\
\hline
trust-constr & 13.1724 \\
\hline
Adadelta & 13.4828 \\
\hline
CG & 15.7241 \\
\hline
TNC & 15.8621 \\
\hline
SLSQP & 16.5172 \\
\hline
BFGS & 16.6207 \\
\hline
L-BFGS-B & 16.7931 \\
\hline
\end{tabular} \caption{Average rank across all tasks for all optimizers} \label{t:tank}
\end{table}  

%\begin{figure}[h]
%    \centering
%    \hspace*{-2cm}
%    \includegraphics[scale=0.4]{loss_diff.png}
%    \caption{Average Loss Difference of Optimizers}
%    \label{fig:ld}
%\end{figure}

\begin{table}[h]
\centering
\begin{tabular}{|l|c|}
\hline
Optimizers & Average Error Distance  \\
\hline
Adam & 0.2453 \\
\hline
AMSGrad & 0.2755 \\
\hline
Nadam & 0.3637 \\
\hline
Adamax & 0.3649 \\
\hline
Adagrad & 0.4498 \\
\hline
RMSProp & 0.525 \\
\hline
Ftrl & 0.6386 \\
\hline
SGD & 0.825 \\
\hline
Powell & 1.4188 \\
\hline
SPSA & 1.487 \\
\hline
COBYLA & 1.5067 \\
\hline
Nelder-Mead & 2.5761 \\
\hline
Adadelta & 2.9463 \\
\hline
trust-constr & 3.3298 \\
\hline
TNC & 3.807 \\
\hline
BFGS & 4.284 \\
\hline
CG & 4.4403 \\
\hline
L-BFGS-B & 4.5956 \\
\hline
SLSQP & 4.8432 \\
\hline
\end{tabular} \caption{Average normalized loss difference across all tasks for all optimizers} \label{t:loss}
\end{table}  

In Table \ref{t:big}, the the same normalized loss difference function is plotted for the large VQE simulations. In this case, only the parameter shift gradient optimizers (which could be efficiently calculated not using the parameter shift, but adjoint differentiation) were compared. In this case, we see that Adam remains dominant with an average of 0 loss difference indicating that it is the best performing in both simulations. We see a similar performance pattern for the other optimizers as well, with the Adam variants coming in just behind Adam.

%\begin{figure}[H]
%    \centering
%    \hspace*{-2cm}
%    \includegraphics[scale=0.4]{big.png}
%    \caption{Average Loss Difference of Optimizers for Large VQE Simulations}
%    \label{fig:big}
%\end{figure}

\begin{table}[h]
\centering
\begin{tabular}{|l|c|}
\hline
Optimizers & Loss Difference  \\
\hline
Adam & 0.0 \\
\hline
AMSGrad & 0.1748 \\
\hline
Nadam & 0.2234 \\
\hline
RMSProp & 0.308 \\
\hline
Adamax & 0.3211 \\
\hline
Adagrad & 0.8255 \\
\hline
Adadelta & 0.9996 \\
\hline
SGD & 0.9997 \\
\hline
Ftrl & 1.0 \\
\hline
\end{tabular} \caption{Performance on Large VQE} \label{t:big}
\end{table}

\section{\label{dis}Discussion}

These results are indicative that two folk-wisdoms about training QVCs are generally true. First, SPSA is an efficient optimizer and effective in noisy situations. Part of the reason SPSA is chosen, especially for real hardware \citep{kandala2017hardware}, is the optimization takes $O(1)$ forward passes for each iteration. This is in contrast to the parameter shift, which takes $O(|\theta|)$ forward passes. SPSA outperforms other black-box optimizers, even those with far more function evaluations required. Additionally, the usage of higher learning rates than classical machine learning is affirmed. The higher learning rates consistently outperform the lower learning rates, even when the lower learning rates have sufficient convergence time. Although this is empirically justified, the exact theoretical reason for this remains unknown. The biggest detour from common understanding that these results show is the potential use of Nelder-Mead. Even in the presence of noise, this algorithm can (although not consistently) perform well. It is for this reason that we recommended an evaluation of Nelder-Mead in our introductory remarks. Given that it can outperform all other methods, it is worth at least some consideration in many applications. The empirical performance warrants further explorations and improvements for the application of Nelder-Mead to QVC optimization.

\subsection{Limitations and Future Work}

\begin{figure}
    \centering
    \includegraphics[scale=0.3]{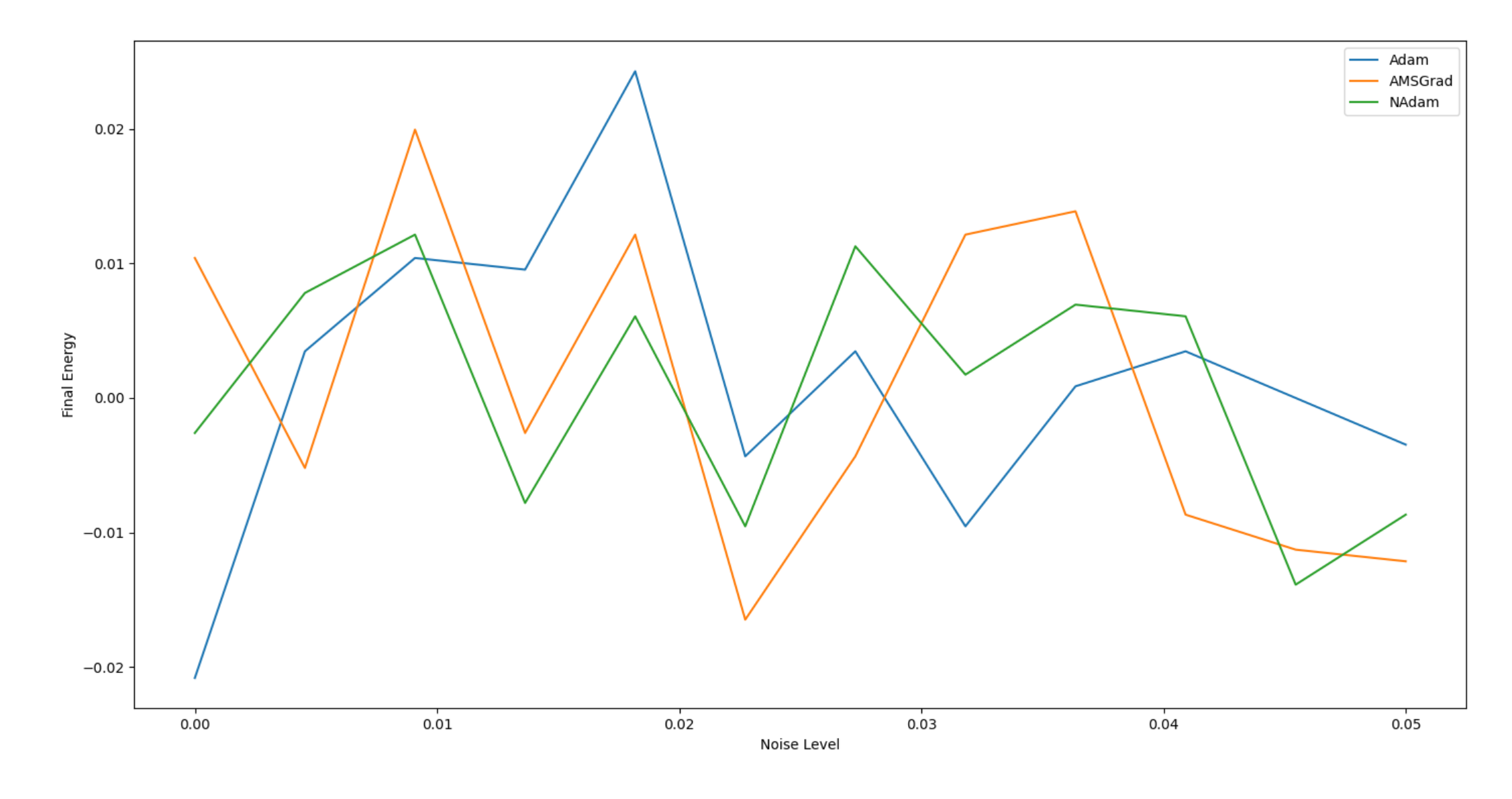}
    \caption{Optimizer performance as a function of noise levels}
    \label{fig:noise}
\end{figure}

The first limitation is our consideration of noise. Our noise modelling (and very low noise level) resulted in almost no differences in the optimization results. Although noise levels can result in substantial variations in optimizer performance, our noise modelling did not. For a concrete example of this, see Figure \ref{fig:noise}. Here we can see the three top performing gradient optimizers and their final results on a simple VQE problem as a function of noise level. This shows that at depolarizing noise levels $<0.05$ this noise has limited impact. However, this claim likely warrants further investigations and the effect of noise (especially real hardware noise) on different optimizers is something that should be expanded upon. 

A second limitation is the consideration (or lack there of) of shot evaluations. A common challenge for users of hardware is the lack of infinite (or very many) shots. We solely consider the performance of the algorithms on a given budget of 1000 optimization steps. However, this is not an entirely fair comparison since some algorithms require substantially fewer shots to complete. E.g. on the excited state classification problem, any parameter shift based method requires 128,000,000 shots for all iterations, where as SPSA requires only 2,000,000 shots. The reason for this choice to ignore shot considerations is primarily due to the forward looking nature of this review. As more quantum hardware (even NISQ hardware) gets produced, the cost per shot will likely dramatically decrease. However, the problems of how to get the best results will remain. 

Although the scope of this review was large, there is room for relevant and important expansions. One such example would be to improve the hamiltonian choice for the VQE simulations. Currently the hamiltonians are randomly generated, but the results for VQE optimization would be more informative if these simulations were actually reflective of the use cases (e.g. if they were evaluated on molecular hamiltonians). This also applies to QAOA. Although MAX-CUT is a common application, there are other optimization problems that QAOA is commonly applied to that could be considered. Circuit structure expansions could also be considered. With a focus on just QCNN and HEA based ansatze, there is substantial room for more investigations e.g. QGNN \citep{verdon2019quantum} or $S_n$-CQA \citep{zheng2021speeding}. A final expansion that should be considered is running on real hardware. Real hardware noise is noticeably different from simulated depolarizing noise \citep{isakov2021simulations} and is a prerequisite for more informative conclusions. However, this work cannot be trivially applied to real hardware, as even a single optimization routine of the excited state classified could cost $>$1,000,000 USD\footnote{\href{https://aws.amazon.com/braket/pricing/}{Price per shot information}}.

\section{\label{conc}Conclusion}

We present the largest scale evaluation of optimization techniques for quantum machine learning methods to date. Using a variety classical and quantum specific optimizers, and we provide empirically justified guidance for future research in the optimization of quantum machine learning systems. This work serves as important guidance for decisions that currently lack strong theoretical or empirical justifications. We reaffirm common choices of parameter shift gradient based optimizers and learning rates, non parameter shift gradient based optimizers. Finally, we lay important groundwork for future expansions and reviews.

\section{Acknowledgements}

This work was funded by the Spell Open Research Grant. 

\newpage 
\clearpage

\bibliographystyle{unsrtnat}
\bibliography{apssamp}

\newpage 

\appendix

\section{\label{add}Additional Results}

\begin{table}[h]
\centering
\begin{tabular}{|l|c|c|}
\hline
Problem & Qubits & Parameters  \\
\hline
VQE & 5 & 40 \\
\hline
VQE & 10 & 80 \\
\hline
QAOA p = 1, 5 Node Graph & 5 & 2 \\
\hline
QAOA p = 1, 10 Node Graph & 10 & 2 \\
\hline
QAOA p = 1, 15 Node Graph & 15 & 2 \\
\hline
QAOA p = 4, 5 Node Graph & 5 & 8\\
\hline
Excited State Classification & 8 & 64 \\
\hline
Moons Classification Depth = 2 & 2 & 8 \\
\hline
Moons Classification Depth = 8 & 2 & 32 \\
\hline
Moons Classification Depth = 16 & 2 & 64 \\
\hline
Regression & 13 & 52 \\
\hline
Blobs Classification & 7 & 28 \\
\hline
Circles Classification Depth = 2 & 2 & 8 \\
\hline
Circles Classification Depth = 8 & 2 & 32 \\
\hline
Circles Classification Depth = 16 & 2 & 64 \\
\hline
%Noisy Circles Classification Depth = 2 & 2 & 8 \\
%\hline
%Noisy Circles Classification Depth = 8 & 2 & 32 \\
%\hline
%Noisy Circles Classification Depth = 16 & 2 & 64 \\
%\hline
%Noisy QAOA p = 1, 5 Node Graph & 5 & 2 \\
%\hline
%Noisy QAOA p = 1, 10 Node Graph & 10 & 2 \\
%\hline
%Noisy QAOA p = 1, 15 Node Graph & 15 & 2 \\
%\hline
%Noisy Blob Classification & 7 & 28 \\
%\hline
%Noisy Regression & 13 & 52 \\
%\hline
%Noisy VQE & 5 & 40 \\
%\hline
%Noisy VQE & 10 & 80 \\
%\hline
%Noisy Excited State Classification & 8 & 64 \\
%\hline
%Noisy Moons Classification Depth = 2 & 2 & 8 \\
%\hline
%Noisy Moons Classification Depth = 8 & 2 & 32 \\
%\hline
%Noisy Moons Classification Depth = 16 & 2 & 64 \\
%\hline
VQE (no shot noise) & 20 & 160 \\
\hline
VQE (no shot noise) & 25 & 200 \\
\hline
\end{tabular} \caption{Problem Sizes} \label{tab:exp1}
\end{table}  

%\begin{figure}[h]
%    \centering
%    \hspace*{-3cm}
%    \includegraphics[scale=0.5]{all_rank.png}
%    \caption{Average Rank of All Optimizers}
%    \label{fig:all_rank}
%\end{figure}

\begin{table}[h]
\centering
\begin{tabular}{|l|c|}
\hline
Optimizer & Average Rank \\
\hline
Powell & 7.1379 \\
\hline
Nelder-Mead & 7.5172 \\
\hline
Nadam 0.01 & 7.5517 \\
\hline
COBYLA & 8.5172 \\
\hline
Ftrl 0.1 & 9.5517 \\
\hline
AMSGrad 0.1 & 9.6552 \\
\hline
Adam 0.1 & 9.8966 \\
\hline
AMSGrad 0.01 & 10.1034 \\
\hline
Adagrad 0.1 & 10.4828 \\
\hline
Adam 0.01 & 10.8621 \\
\hline
Nadam 0.1 & 11.5517 \\
\hline
RMSProp 0.1 & 12.0345 \\
\hline
Adamax 0.1 & 12.7241 \\
\hline
RMSProp 0.01 & 14.0 \\
\hline
SPSA & 14.6552 \\
\hline
Adamax 0.01 & 15.4483 \\
\hline
Nadam 0.001 & 17.2759 \\
\hline
AMSGrad 0.001 & 20.1034 \\
\hline
Ftrl 0.01 & 20.5172 \\
\hline
Adam 0.001 & 21.7931 \\
\hline
SGD 0.1 & 22.3103 \\
\hline
Adagrad 0.01 & 24.2759 \\
\hline
trust-constr & 24.931 \\
\hline
SGD 0.01 & 25.5172 \\
\hline
RMSProp 0.001 & 25.7931 \\
\hline
SGD 0.001 & 28.7586 \\
\hline
SGD 0.0001 & 29.9828 \\
\hline
Adamax 0.001 & 30.3103 \\
\hline
Ftrl 0.0001 & 30.9655 \\
\hline
Ftrl 0.001 & 32.1034 \\
\hline
AMSGrad 0.0001 & 32.8966 \\
\hline
CG & 33.1379 \\
\hline
TNC & 33.3448 \\
\hline
Adam 0.0001 & 33.3448 \\
\hline
Nadam 0.0001 & 33.3448 \\
\hline
Adadelta 0.01 & 33.5862 \\
\hline
Adadelta 0.1 & 33.9655 \\
\hline
Adagrad 0.001 & 34.3103 \\
\hline
RMSProp 0.0001 & 35.431 \\
\hline
Adagrad 0.0001 & 35.5517 \\
\hline
Adadelta 0.0001 & 35.6207 \\
\hline
SLSQP & 35.6552 \\
\hline
Adamax 0.0001 & 35.7241 \\
\hline
L-BFGS-B & 36.069 \\
\hline
BFGS & 36.2759 \\
\hline
Adadelta 0.001 & 36.4138 \\
\hline
\end{tabular} \caption{Average rank across all tasks} \label{tab:all_rank}
\end{table}

%\begin{figure*}[h]
%    \centering
%    \hspace*{-3cm}
%    \includegraphics[scale=0.5]{all_loss_diff.png}
%    \caption{Average Loss Difference of All Optimizers}
%    \label{fig:ald}
%\end{figure*}
\begin{table}[h]
\centering
\begin{tabular}{|l|c|}
\hline
Optimizer & Average Normalized Error Difference \\
\hline
AMSGrad 0.1 & 0.347 \\
\hline
Adam 0.1 & 0.4149 \\
\hline
Nadam 0.01 & 0.4351 \\
\hline
Adagrad 0.1 & 0.4605 \\
\hline
Adamax 0.1 & 0.4869 \\
\hline
Adam 0.01 & 0.5097 \\
\hline
RMSProp 0.1 & 0.5913 \\
\hline
Ftrl 0.1 & 0.6386 \\
\hline
AMSGrad 0.01 & 0.6771 \\
\hline
Nadam 0.1 & 0.9747 \\
\hline
RMSProp 0.01 & 1.0595 \\
\hline
Adamax 0.01 & 1.3931 \\
\hline
Powell & 1.4188 \\
\hline
SGD 0.1 & 1.4816 \\
\hline
SPSA & 1.487 \\
\hline
COBYLA & 1.5067 \\
\hline
SGD 0.01 & 2.0864 \\
\hline
AMSGrad 0.001 & 2.0961 \\
\hline
Nadam 0.001 & 2.2122 \\
\hline
Adam 0.001 & 2.3354 \\
\hline
Ftrl 0.01 & 2.4637 \\
\hline
Nelder-Mead & 2.5761 \\
\hline
Adagrad 0.01 & 2.8652 \\
\hline
SGD 0.001 & 3.1458 \\
\hline
trust-constr & 3.3298 \\
\hline
RMSProp 0.001 & 3.3933 \\
\hline
SGD 0.0001 & 3.4975 \\
\hline
Adamax 0.001 & 3.7249 \\
\hline
Adagrad 0.001 & 3.7719 \\
\hline
Adam 0.0001 & 3.7732 \\
\hline
TNC & 3.807 \\
\hline
Adamax 0.0001 & 3.8156 \\
\hline
Adadelta 0.1 & 3.8459 \\
\hline
Adadelta 0.01 & 3.9861 \\
\hline
Nadam 0.0001 & 4.0573 \\
\hline
Adadelta 0.0001 & 4.145 \\
\hline
Adagrad 0.0001 & 4.2499 \\
\hline
BFGS & 4.284 \\
\hline
Adadelta 0.001 & 4.2963 \\
\hline
Ftrl 0.001 & 4.3596 \\
\hline
CG & 4.4403 \\
\hline
AMSGrad 0.0001 & 4.4544 \\
\hline
RMSProp 0.0001 & 4.581 \\
\hline
L-BFGS-B & 4.5956 \\
\hline
SLSQP & 4.8432 \\
\hline
Ftrl 0.0001 & 4.8561 \\
\hline
\end{tabular} \caption{Average normalized loss difference across all tasks} \label{tab:all_loss}
\end{table}  

%\nocite{*}

\end{document}